\documentclass[12pt]{article}
\usepackage[dvips]{graphicx}
\def\be{\begin{equation}}					 
\def\ee{\end{equation}}
\def\ber{\begin{eqnarray}}
\def\eer{\end{eqnarray}}	
\def\dint{\mathop{\intop\kern-0.5em\intop}}
\def\ovc#1{\displaystyle\mathop{#1}^{\kern0.2em\circ}}

\begin{document}

\begin{center}
{\large \bf Harmony of the Froissart Theorem \\[1ex]in Fundamental
Dynamics of Particles and Nuclei}\\

\vspace{4mm}

A.A. Arkhipov\\
{\it State Research Center ``Institute for High Energy Physics" \\
 142280 Protvino, Moscow Region, Russia}\\
\end{center}

\begin{abstract}
It has been shown that the great ancient Pythagorean ideas have found
themselves in the latest researches in high energy elementary
particles and nuclear physics. In this respect we concern and
discuss the mathematical, physical and geometrical aspects of the
famous Froissart theorem and in this way one establishes a link of
this theorem to the mathematics and ideas elaborated in the
Pythagorean school. A harmony of the Froissart theorem in fundamental
dynamics of particles and nuclei has been displayed. We argue that a
harmony of the Froissart theorem allow us to hear the new notes
of {\it ``the music of the spheres"} just in the Pythagoreans sense. 
\end{abstract}

\vspace{4mm}
\hfill{\it ``The Master said so"}

\hfill Pythagorean watchword 

\section*{Introduction: Pythagoreanism}
\font\cyr wncyr10 at 12.0pt
\hfill
\begin{minipage}{103mm}
{\bf 11.} ... \rm{\cyr za nerazumnye pomyshleniya ih nepravdy,
po kotorym oni sluzhili besslovesnym presmykayuwimsya i prezrennym
chudoviwam, Ty v na\-ka\-zanie poslal na nih mnozhestvo besslovesnyh
zhivotnyh, chtoby oni poznali, chto chem kto sogreshit, tem i
nakazyvaet\-sya ... i bez \char"0Btogo oni mogli pogibnut\char"7E\ \
ot
odnogo dunoveniya, presledu\-emye pravosudiem i rasseivaemye duhom
sily
tvoe\char"1A; no \\
Ty vse raspolozhil meroyu, chislom i vesom}.
\font\cyrb wncyb10 at 12.0pt

\hfill {\cyrb Kniga premudrosti Solomona}

\end{minipage}

\vspace{1cm}
\noindent
Once upon a time the great russian physicist-theorist and
mathematician Bogoljubov said that the last line in the above written
fragment from the Book of Proverbs which is a part of the
Bible (non-canonical though!) ``...; but You did all arrange  by
measure, number and weight" represents the definition of the Physics
\cite{1}. Probably this -- ``...; but You did all arrange  by
measure, number and weight" -- was an earliest evidence of the
principle that All in the World have to be in harmony with
each other. 

In fact,
the word harmony (Syn: music: accord, concord, consonance) has the
Greek origin from $\check\alpha\rho\mu\acute o\nu\acute\iota\alpha$
which means orderliness (symmetry) of the whole, commensurability
(proportionality) of its parts. The idea of harmony has intensively
been elaborated by Pythagoras, the Greek philosopher and
mathematician and founder of the Pythagorean school \cite{2}.
Originally from
Samos, Pythagoras founded  a society which was at once a religious
community and a scientific school flourished at Kroton in
Southern Italy about the year 530 B.C. Pythagoras was the first
genius of western culture. He had a multifaceted magnetic personality
-- an intelligent mathematician and a religious thinker, both
co-existed in him. His main contributions are in geometry, numbers,
music, cosmology, astronomy, philosophy and religion. Pythagoras must
have been one of the world's greatest men, but he wrote nothing
though numerous
works are attributed to him, and it is hard to
say how much of the doctrine one knows as Pythagorean is due to the
founder of the society and how much is later development. It is also
hard to say how much of what we are told about the life of Pythagoras
is trustworthy. For a mass of legend gathered around his name:
Sometimes he is represented as a man of science, and
sometimes as a preacher of mystic doctrines, and we might be tempted
to regard one or other of those characters as alone historical.
Certainly, it's true that there is no need to reject either of the
traditional
views.

Even though many wonderful
things related to Pythagoras, belong to legend, and seem to have no
historical foundation, similarly the description of the learned works
which he wrote is not
attested by reliable historians and  also belongs to the region of
fable, nevertheless it is no doubt however, that he founded a school,
or, rather, a religious philosophical society, which exerted great
influence on the intellectual development of human civilization and
had a fundamental importance all the time. Of great influence were
the Pythagorean doctrines that
numbers were the basis of all things and possessed a mystic
significance, in particular the idea that the cosmos is a
mathematically ordered whole. Aristotle wrote: ``Pythagorean having
been brought up in the study of mathematics, thought that things
could be represented by numbers ... and that the whole cosmos
consists of a scale and a number". Briefly stated, the doctrine of
Pythagoras was that all things are numbers. Pythagoras was led to
this conception by his discovery that the notes sounded  by stringed
instrument are related to the length of the strings. He 
conducted remarkable investigation in ``music" as he was a
musician. Harmonies correspond to most beautiful mathematical ratio,
he stated. Melodious musical tunes could be produced on a stringed
instrument by plucking the string at particular points, which
correspond to mathematical ratios. Such beautiful mathematical ratios
are $1:2$ (an octave), $2:3$ (a fifth), and $3:4$ (a fourth).
Pythagoras recognized that first four numbers $1,2,3,4$ known as
``tetractys", whose sum equals Ten  
($1+2+3+4=10$), contained all basic musical intervals: the octave,
the
fifth and the fourth. In fact, all the major consonances, that is,
the octave, the fifth and the fourth are produced by vibrating
strings whose lengths stand to one another in the ratios of $1:2$,
$2:3$ and $3:4$ respectively. Recent major scale in according to
Pythagoras tune looks like

\[
1; \frac{8}{9}; \frac{64}{81}; \frac{3}{4}; \frac{2}{3};
\frac{16}{27}; \frac{128}{243}; \frac{1}{2},
\]
where $\frac{8}{9}=\frac{2}{3}\cdot\frac{2}{3}\cdot 2$ is major
second (fifth of fifth with octave lowering);
$\frac{16}{27}=\frac{2}{3}\cdot\frac{8}{9}$ is major sixth (fifth of
major second);  $\frac{64}{81}=\frac{2}{3}\cdot\frac{16}{27}\cdot 2$
is major third (fifth of sixth with octave lowering); 
$\frac{128}{243}=\frac{2}{3}\cdot\frac{64}{81}$ is major seventh
(fifth of third).

The resemblance which Pythagoras perceived
between the orderliness of music, as expressed in the ratios which he
had discovered and the idea that cosmos is an orderly whole, made up
of parts harmoniously related to one another, led him to conceive of
the cosmos too as mathematically ordered. Pythagoras compared the
eight planets (there were seven planets known the Babylonians: Moon,
Mercury, Venus, Sun, Mars, Jupiter and Saturn), including the Earth,
with the musical octave and the seven planets, excluding the Earth,
as seven strings of the musical instrument Lair. The planets situated
at different distances and moving at different speed correspond to
different notes on musical octave. The planets moving with higher
speed produce higher notes and those with lower speed produce lower
notes. The celestial harmony of moving planets produces heavenly
music (``the music of the spheres") analogous to different notes of
musical octave. 
         
According to Pythagoras, the sphere was the most beautiful solid and
the circle the most beautiful shape. Thus, a spherical planet moving
in circular orbit would form a harmonious constellation. Pythagoras
worked out the distances of the planets
from the Earth. He arranged the planets in order of increasing
distances of the planets from the Earth. The order given by him was
the Moon, Mercury, Venus, the Sun, Mars, Jupiter and Saturn. Some
Pythagoreans believed that the Earth moved round a central fire. The
Earth did not always face the central fire.	This accounted for day
and night on the Earth. They also believed
that the Moon as well as the Sun shone because they reflected light
from their surfaces received from the central fire. Perhaps the idea
of central fire later on led to the heliocentric (Sun at the centre
of the Solar system) configuration of Solar system. Pythagoras
observation of heavens suggested to him that the motion
of the heavenly bodies was cyclic and that the heavenly bodies
returned to the place from which they had started. From this,
Pythagoras concluded that there must be a cycle of cycles, a greater
year and on its completion the heavenly bodies returned to the
original position and the same heavenly constellation would be
observed again and again. He called this the eternal recurrence.

Pythagoras doctrine that mathematics contains the key to all
philosophical  knowledge, an idea, which was by his followers
afterwards developed into an elegant number-theory. 
The Pythagorean philosophy in its later  elaboration is dominated by
the number-theory. Being the first, apparently, to observe that
natural
phenomena, especially the phenomena of the astronomical world, may be
expressed in  mathematical formulas, the Pythagoreans held that
numbers  are not only the symbols of reality, but the very substance
of real things. Pythagoras associated numbers with geometrical
notions and numerical ratios with shapes. He associated number one
with a point, too with a line, three with a triangle (the surface)
and four with a tetrahedron (the solid). Thus, one point generates
dimensions, two points generate a line of
one dimension, three points generate a surface of two dimensions, and
four points generate three-dimensional solid figures. In geometry,
numbers represent lengths, their squares represent areas, their cubes
represent volumes. Starting from numbers, numerical ratios and
their powers, one can construct geometrical figures of different
shapes and geometrical solids of different sizes. Using distance the
arrangement of planets, their motion, their orbital path, their
distances from the center and their interrelations with each other
can be worked out. Thus, according to Pythagoras all relations could
be reduced to number relations and hence, the whole cosmos is a scale
and a number based phenomenon.

According to Pythagoras,  Ten is the perfect number, because it is
the sum of  one, two, three, and four -- the point, the line, the
surface, and the  solid. There are the second type of perfect numbers
: According to Pythagoras the second type of perfect numbers are
those were the numbers equal to sum of their factors. For instance 28
has factors 1, 2, 4, 7, 14  and $1+2+4+7+14=28$.

From perfect numbers, Pythagoras was led to amicable numbers like 220
and 284. Amicable numbers form a pair of numbers where each number is
equal to the sum of the factors of the other numbers. For instance
220 has factors 1, 2, 4, 5, 10, 11, 20, 22, 44, 55, 110. The sum of
these factors is $1+2+4+5+10+11+20+22+44+55+110=284$. Moreover, 284
has factors 1, 2, 4, 71 , 142 . The sum of these factors is
$1+2+4+71+142=220$.

Triangular numbers have been introduced by Pythagoras: Pythagoras
called numbers 1, 3, 6, 10, 15, 21, 28, 36, 45, 55, 66 as
triangular numbers because these numbers can be arranged so as to
form triangles.

If $a,b,c$ are sides of a right-angled triangle and $c$ is the
hypotenuse then according to Pythagoras theorem $c^2 = a^2 +
b^2$. The triad of positive integers $(a,b,c)$ satisfying the
relation $c^2 = a^2 + b^2$ is called the Pythagorean triad of
numbers. About fifteen such triads were previously known like
(3,4,5), (5,12,13), (7,24,25), (9,12,15), (15,36,39). The Pythagorean
triads in which the numbers $a, b, c$ do not have a common factor are
called primitive Pythagorean triads. For example (3,4,5), (5,12,13),
(7,24,25) etc. are primitive Pythagorean triads. But (9,12,15),
(15,36,39) are not primitive triads. It is believed
that Pythagoras himself discovered the formula for determining triads
of numbers satisfying the relation $c^2 = a^2 + b^2$. In fact, all
Pythagorean triads can be expressed via formulae
\be
a=m^2-n^2,\quad b=2mn,\quad c=m^2+n^2,\label{triad}
\ee
where $m,n$ are any positive integers $(m>n>0)$.

From his observations in music, mathematics and astronomy, Pythagoras
generalized that everything could be expressed in terms of numbers
and numerical ratios. Numbers are not only symbols of reality, but
also substances of real things. Hence, he claimed - All is number.
The importance of this conception is very great, for example, it is
the ultimate source of Galileo's belief {\it
``Il libro della natura \'e scritto in lingua matematica"}
that the book of nature is written in mathematical symbols and
hence the ultimate source of modern physics in the form in which it
came to us from Galileo. 

It may be taken as certain that the union of
mathematical genius and mysticism is common enough\footnote{One
up-to-date outstanding mathematician contended that all scientists,
working in the number-theory, have a conversation with the God.}. 
Pythagoras himself discovered the numerical ratios which determine
the concordant intervals of the musical scale. Similar to musical
intervals, in medicine there are opposites, such as the hot and the
cold, the wet and the dry, and it is the business of the physician to
produce a proper ``blend" of these in the human body. The
Pythagoreans contended that the opposites are  found everywhere in
Nature, and the union of them constitutes the  harmony of the real
world. They also argued for the  notion that virtue is a harmony, and
may be cultivated not only by  contemplation and meditation but also
by the practice of  gymnastics and music. 

Pythagoras held the theory that what gives form to the Unlimited is
the Limit. That is the great contribution of Pythagoras to
philosophy, and we must try to understand it. It was natural for
Pythagoras to look for something of the same kind in the world at
large. Musical tuning and health are alike means arising from the
application of Limit to the Unlimited. 

In  their psychology and their ethics the Pythagoreans used the idea
of  harmony and the notion of
number as the explanation of the mind and  its states, and also of
virtue and its various kinds. Pythagoras argued that there are three
kinds of men, just as there are three classes of strangers who come
to the Olympic Games. The lowest consists of those who come to buy
and sell, and next above them are those who come to compete. Best of
all are those who simply come to look on. Men may be classified
accordingly as lovers of wisdom, lovers of honour, and lovers of
gain.
That seems to imply the doctrine of the tripartite soul, which is
also attributed to the early Pythagoreans on good authority.

The Pythagoreans were  religiously and ethically inclined,
and strove to bring philosophy  into relation with life as well as
with knowledge. The Pythagoreans believed also in
reincarnation or transmigration (doctrine of Rebirth), that is, the
soul, after death, passes into another
living thing, which presupposes the ability of the soul to survive
the death of the body, and hence some sort of belief in its
immortality. 

The above detailed introduction is made so as to show in the
next sections that the great ancient Pythagorean ideas have found
themselves in the latest researches in high energy elementary
particle and nuclear physics. In this respect we will concern and
discuss the mathematical, physical and geometrical aspects of the
famous Froissart theorem and in this way we will easily establish a
link of this theorem to the mathematics and ideas elaborated in the
Pythagorean school. In other words, we would like to show a harmony
of the Froissart theorem just in the Pythagoreans sense. 

\section{Froissart theorem: mathematical, physical and geometrical
aspects}

In the year 1961 french physicist Marcel Froissart discovered and
proved a remarkable theorem, which stated that two-body reaction $a +
b \rightarrow c + d$ amplitude, satisfying Mandelstam representation,
is bounded by expressions of the form $C s\, {\ln^2}s$ at the forward
and backward angles, and $C s^{\frac{3}{4}}{\ln^\frac{3}{2}}s$ at any
fixed angle in the physical region, $C$ being a constant, $s$ being
the total squared c.m. energy (one of the Mandelstam invariant
variables $s, t, u$). This corresponds to the total cross sections
increasing at most like ${\ln^2}s$ \cite{3}. A little bit later it
was shown  that the analytical properties of two-particle scattering
amplitude, which may be established strictly in the framework of
axiomatic Quantum Field Theory, bring us to the Froissart statements
as well. Up-to-date derivation of the Froissart theorem can be
realized in a few steps, and we briefly sketch out it here. 

For simplicity we consider a reaction of elastic scattering $a + b
\rightarrow a + b$ for two scalar particles. The scattering amplitude
of the two-body reaction may be considered as a function of the
invariant variable $s=(p_a+p_b)^2$ and two unit vectors ${\bf n}$ and
${\bf n}'$ on two-dimensional sphere $S_2$, which characterise the
initial and final states of two-particle system: ${\cal F}_2(s;p'_a
p'_b,p_a
p_b)={\cal F}_2(s;{\bf n}',\bf n), {\bf n}={\bf q}/|{\bf q}|$, ${\bf
q}$ is
c.m. momentum of particles in an initial state 
\[
{\bf q}={\ovc{\bf p}}_a=-{\ovc{\bf p}}_b,\quad  {\ovc{\bf p}}_{a,b}
=
\stackrel{\longrightarrow}{L^{-1}(P_{ab})p_{a,b}},\quad
P_{ab}=p_a+p_b,
\]
$L(P)$ is Lorentz boost, and the same with the primes in a final
state. In the first step we wright the partial wave expansion
\[
{\cal F}_2(s;{\bf n}',{\bf n})={\cal F}_2(s;{\bf n}'\cdot{\bf n}) =
{\cal F}_2(s;\cos\theta)= 
\]
\be
=\frac{1}{\pi A_2(s)}\sum_{lm}Y _{lm}
({\bf n}')f_{l}(s)\stackrel{*}{Y}_{lm}({\bf n})
=\frac{1}{\pi \Gamma_2(s)}\sum_{l}(2l+1)f_l(s)P_l({\bf n}'\cdot{\bf
n}),
\label{1}
\ee
where $A_2(s)=\Gamma _2(s)/S_2$, $\Gamma _2(s)$ is two-particle phase
space volume, $S_2$ is a surface of two-dimensional unit sphere,
$\cos\theta={\bf n}'\cdot{\bf n}$, and  an addition theorem for the
spherical harmonics in second line of Eq. (\ref{1}) has been used.
The second invariant Mandelstam variable $t$ (momentum transfer) is
related to $\cos\theta$ by the following Equation
\be
\cos\theta = 1 + \frac{t}{2{\bf q}^2}.\label{2}
\ee

A remarkable analytic properties of scattering amplitudes as
functions of momentum transfer have been discovered in the year 1958
by Harry Lehmann \cite{4} using Jost-Lehman-Dyson representation
especially Dyson's theorem for a representation of causal commutators
in local Quantum Field Theory \cite{5,6,7}. Lehmann proved that
imaginary part of two-body interaction amplitude
is analytic function of $\cos\theta$, regular inside an ellipse in
$\cos\theta$-plane with center at the origin and with semi-major axis
\be
z_0(s) = 1 + \epsilon_L(s), \quad \epsilon_L(s) =
\frac{2(m_1^2-m_a^2)(m_2^2-m_b^2)}{{\bf
q}^2[s-(m_1-m_2)^2]},\label{3}
\ee
where $m_1$ and $m_2$ define the support of spectral function in the
JLD representation by the requirements of spectral condition or
spectrality. Actually, $m_1$ and $m_2$ are the lowest mass values of
the physical states for which the following matrix elements are not
equal to zero
\[
<0|J_a(0)|m_1> \not= 0,\quad <0|J_b(0)|m_2> \not= 0,
\]
where $J_a(x)$ and $J_b(x)$ are local Heisenberg's currents of
particles $a$ and $b$.
He also shown that two-body interaction amplitude, as
itself, is analytic function of $\cos\theta$, regular inside an
ellipse in $\cos\theta$-plane with center at the origin and with
semi-major axis $x_0(s)$ which is related to $z_0(s)$ by the Equation
\be
x_0(s) = \sqrt{\frac{z_0(s)+1 }{2}}.\label{4}
\ee
Afterwards the fundamental results of Harry Lehmann were improved by
Martin \cite{8} and Sommer \cite{9}: it was shown that imaginary part
of two-body interaction amplitude is analytic function of
$\cos\theta$, regular inside an ellipse in $\cos\theta$-plane with
semi-major axis
\be
z_0(s) = 1 + \epsilon_M(s), \quad \epsilon_M(s) =
\frac{t_0}{2{\bf q}^2}, \quad t_0 = 4{m_\pi}^2,\label{5}
\ee
\be
{\bf q}^2 = \frac{\lambda(s,m_a^2,m_b^2)}{4s} =
\frac{[s-(m_a+m_b)^2][s-(m_a-m_b)^2]}{4s},\label{lambda}
\ee
where ${m_\pi}$ is pion mass. Correspondingly two-body interaction
amplitude, as itself, appears as analytic function of $\cos\theta$,
regular inside an ellipse in $\cos\theta$-plane with semi-major axis
$x_0(s)$ which is related to $z_0(s)$
by Eq. (\ref{4}).

The fundamental results derived by Lehmann and improved by his
followers are of great importance because it has been shown that the
partial wave expansions (\ref{1}) which define physical scattering
amplitudes continue to converge for complex values of the scattering
angle, and define uniquely the amplitudes appearing in the unphysical
region of non-forward dispersion relations. In fact, expansions
converge for all values of momentum transfer for which dispersion
relations have been proved. The proved analyticity of two-body
interaction amplitudes as functions of two complex Mandelstam
variables $s$ and $t$ in a topological product of cut 
$s$-plane with the cuts ($s_{thr}\leq s \leq\infty,\,  u_{thr}\leq
u \leq\infty$) except for possible fixed poles and circle
$|t|\leq t_0$ in $t$-plane allowed in a more general case to save the
fundamental Froissart results previously obtained at a more
restricted Mandelstam analyticity.
Really, let us wright Cauchy representation for imaginary part of 
two-body interaction amplitude
\[
Im {\cal F}_2(s;\cos\theta)=\frac{1}{2\pi i}\oint_C dz \frac{Im
{\cal F}_2(s;z)}{z
- \cos\theta},
\]
where contour $C$ is a boundary of an ellipse in $\cos\theta$-plane
with semi-major axis given by Eq. (\ref{5}). Using Heine formula
\[
\frac{1}{z - \cos\theta} =
\sum_{l=0}^{\infty}(2l+1)Q_l(z)P_l(\cos\theta),
\]
we obtain
\be
Im f_l(s) = \frac{\Gamma_2(s)}{2 i}\oint_C dz Im
{\cal F}_2(s;z)Q_l(z).\label{6}
\ee
From Eq. (\ref{6}) it follows
\be
Im f_l(s)\leq \frac{1}{2}\Gamma_2(s)\cdot\max_{z\in C}|Im
{\cal F}_2(s;z)|\cdot\max_{z\in C}|Q_l(z)|\cdot{\cal L}(C),\label{7}
\ee
where ${\cal L}(C)$ is a length of contour $C$. Representation
(\ref{6}) where estimate (\ref{7}) followed from is a good tool to
study an asymptotic behaviour of partial waves at large orbital
momentum. Using asymptotic properties of the Legendre functions $Q_l$
\cite{10}
\[
Q_l(z)\simeq \sqrt{\frac{\pi}{2l}}(z^2-1)^{-\frac{1}{4}}(z +
\sqrt{z^2-1})^{-l-\frac{1}{2}},\quad
|l|\rightarrow\infty, \quad |\arg l| < \pi,\quad z\in C
\]
and polynomial boundedness $$\max_{z\in C}|Im {\cal F}_2(s;z)|\leq
P_2(s),$$
$P_2(s)$ is some polynomial in $s$, we find
\be
Im f_l(s)\leq
\sqrt{\frac{2\pi}{l}}\Gamma_2(s)P_2(s)\left(\frac{z_0(s)+\sqrt{z_0^2(
s)-1}}{\sqrt{z_0^2(s)-1}}\right)^{\frac{1}{2}}[z_0(s)+\sqrt{z_0^2(s)-
1}]^{-l}, \ \, |l|\rightarrow\infty. \label{8}
\ee
If we put $z_0(s) = 1 + \epsilon(s)$, $\epsilon(s) << 1,\,
s\rightarrow \infty$ then estimate (\ref{8}) at large values of $s$
may be rewritten in the form
\be
Im f_l(s)\leq \frac{{\tilde
P}_2(s)}{\sqrt{l}}\exp\left(-l\sqrt{2\epsilon(s)}\right), \quad
s\rightarrow
\infty,\label{9}
\ee
where
\be
{\tilde P}_2(s) =
\left(\frac{2\pi}{\sqrt{2\epsilon(s)}}\right)^{\frac{1}{2}}\Gamma_2(s
)P_2(s).\label{polinom}
\ee

Thus we have obtained a very important result: analyticity of
two-body interaction amplitudes as functions of $\cos\theta$, regular
inside an ellipse in $\cos\theta$-plane, results in exponential
decrease of partial waves as functions of orbital momentum $l$ at
large values of $l$. This means that the significant contribution to
the partial wave expansion (\ref{1}) is determined by partial waves
for which the orbital momentum does not exceed the quantity
\be
L = \left[\frac{\ln{\tilde
P}_2(s)}{\sqrt{2\epsilon(s)}}\right].\label{10}
\ee
The contribution of partial waves with $l>L$ to the partial wave
expansion will be exponentially small. Let us decompose the partial
wave expansion in two terms
\be
Im {\cal F}_2(s;\cos\theta=1) =
\frac{1}{\pi\Gamma_2(s)}\sum_{l=0}^{L-1}(2l+1)Im f_l(s)
+ Im {\cal F}_2^L(s),\label{11}
\ee
where the second term in Eq. (\ref{11}) contains the contribution of
partial waves with $l\geq L$. Now we would like to take advantage of
unitarity condition which can be written for the partial waves as the
following sequence of inequalities
\be
0\leq |f_l(s)|^2\leq Im f_l(s)\leq |f_l(s)|\leq 1.\label{12}
\ee
Taking into account the unitarity condition we get for the first term
in Eq.~(\ref{11}) an estimate in the form
\be 
\frac{1}{\pi\Gamma_2(s)}\sum_{l=0}^{L-1}(2l+1)Im f_l(s)\leq
\frac{1}{\pi\Gamma_2(s)}\sum_{l=0}^{L-1}(2l+1) = 
\frac{L^2}{\pi\Gamma_2(s)} = \frac{\ln^2{\tilde
P}_2(s)}{2\pi\epsilon(s)\Gamma_2(s)},\label{13}
\ee
where expression (\ref{10}) for the quantity $L$ has been used.

Froissart has shown that the second term in Eq.~(11) is
asymptotically small compared to the first one at large values of
$s$, so that we finally get
\be
Im {\cal F}_2(s;\cos\theta=1) < \frac{\ln^2{\tilde
P}_2(s)}{2\pi\epsilon(s)\Gamma_2(s)}.\label{14}
\ee
The optical theorem relates a total cross section of two-body
interaction with imaginary part of two-body forward elastic
scattering amplitude
\[
\sigma_{ab}^{tot}(s) =
\frac{(2\pi)^3}{\lambda^{1/2}(s,m_a^2,m_b^2)}Im {\cal F}_2
(s;\cos\theta=1),
\]
$\lambda$-function is defined by Eq.~(\ref{lambda}). Hence from
estimate (\ref{14}) it follows an upper bound for the
total cross section of two-body interaction 
\be
\sigma_{ab}^{tot}(s) < \frac{S_2\ln^2{\tilde
P}_2(s)}{32s\epsilon(s)A_2^2(s)}.\label{15}
\ee
where, as it was mentioned above,
\[
A_2(s) = \Gamma_2(s)/S_2 = \frac{\lambda^{1/2}(s,m_a^2,m_b^2)}{8s}.
\]

Here is just the place to introduce the physical notion of the
effective radius of two-body forces \cite{12,13}. Let us define the
effective radius $R_2(s)$ of two-body forces by the following
equation
\be
R_2(s)\stackrel{def}{=}\frac{L}{|\bf q|} = \frac{2\sqrt{s}\ln{\tilde
P}_2(s)}{\sqrt{2\epsilon(s)\lambda(s,m_a^2,m_b^2)}},
\label{16}
\ee
where the definition (\ref{10}) of the quantity $L$ and expression
(\ref{lambda}) for ${\bf q}$ have been used. Now upper bound
(\ref{15}) in terms of such defined quantity $R_2(s)$ takes the form
\be
\sigma_{ab}^{tot}(s)<4\pi R_2^2(s).\label{17}
\ee
This form of the upper bound for experimentally measured quantity
$\sigma_{ab}^{tot}(s)$ has a quite transparent physical and clear
geometrical meanings: it means that the total cross section of
two-body interaction is bounded by the area of a surface of
two-dimensional sphere whose radius is defined by the effective
radius of two-body forces. A remarkable property of upper bound
(\ref{17}) consist in the fact that here all information about
analytic properties of two-body interaction amplitudes is hidden in
the physically tangible quantity (\ref{16}) which is the effective
radius of two-body forces. If we put $\epsilon(s)$ equal to
$\epsilon_M(s)$ given by Eq.~(\ref{5}) then from Eqs.~(\ref{13}) and
(\ref{polinom}) it follows that
\[
{\tilde P}_2(s)\sim {\tilde c}_2\,s^{9/4},\quad s\rightarrow\infty.
\]
In that case for the the effective radius of two-body forces we find
from Eq.~(\ref{16})
\be
R_2(s) = \frac{\ln {\tilde P}_2(s)}{\sqrt{t_0}} \sim 
\frac{9}{4\sqrt{t_0}}\ln (s/s_0) = \frac{9}{8m_\pi}\ln
(s/s_0),\quad s\rightarrow\infty.\label{18}
\ee
In the article Froissart gave an excellent semiclassical explanation
corroborating his theorem. We would like to present here a remarkable
fragment from section II of the Froissart paper \cite{3}. He wrote:
``{\it To get intuitive idea why the amplitude is bounded in the
physical region, let us consider a classical problem: Two particles
interact by means of absorptive Yukawa potential $g\,e^{-\kappa
r}/r$. If $a$ is the impact parameter, the total interaction seen by
a particle for large $a$ is likely to be approximately $g\,e^{-\kappa
a}$. If this is small compared to one, there will be practically no
scattering. If $|g\,e^{-\kappa a}|$ is large compared to one, there
will be practically complete scattering, so that the cross section
will be essentially determined by the value $a=(1/\kappa)\ln|g|$
where $|g\,e^{-\kappa a}|=1$. It is $\sigma \cong
(\pi/\kappa^2)\ln^2|g|$. If we now assume that $g$  is a function of
the energy, and increases like a power of the energy, then $\sigma$
will vary at most like the squared logarithm of the energy}." In
fact, Froissart anticipated here a running coupling and
quasi-potential character of strong forces. Later on it was shown
\cite{14} that the hypothesis about validity of the dispersion
relations in the momentum transfer leads, for any value of the energy
$s$, to a potential which is a superposition of Yukawa potentials
with
energy dependent intensities. This fact together with a theorem on
single-time reduction in Quantum Field Theory \cite{15} provides a
strong basis for semiclassical consideration given by Froissart.
However, it should be stressed that upper bound (\ref{17}) has a
quite different geometrical sense compared to semiclassical
consideration given by Froissart: Eq.~(\ref{17}) shows that the total
cross section of two-body interaction is bounded by the area of a
surface of the sphere with the radius equal to the effective radius
of two-body forces but not by the area of a disk with the same
radius. 

Unitarity bound (\ref{17}) states that the total probability (per
unit volume per unit time in fraction of particles density flux) of
all possible (elastic and inelastic) two-particle interactions, which
take place in a limited volume $V$ during a limited interval of time
$T$, is limited by the area of a surface of the sphere which is,
actually, a boundary of the volume $V$. This means that widely
discussed in the recent literature concerning some physical problems
at Planck scale {\it the holographic principle} \cite{16} has been
incorporated in the general scheme of axiomatic Quantum Field Theory
and resulted from the general principles of local Quantum Field
Theory.    

\section{Generalized Froissart theorem}

In our works \cite{17,18} it was shown that there is a quite natural
geometrical generalization of the Froissart theorem to the case of
multiparticle interaction. In this respect it should be noted that
the problem of finding such generalization is non-trivial because at
least the known singularities of multiparticle scattering amplitudes
related to disconnected parts by cluster structure of the amplitudes
point to the fact that for the total amplitude of $n$-particle
scattering ($n\geq 3$) there is no such generalization. Connected
part of $n$-particle ($n\geq 3$) scattering  amplitudes contains
singular rescattering terms as well. Therefore, the first problem
which arises in this case is to define a suitable object connected
with the $n\rightarrow n$ reaction amplitude which would permit a
correct formulation of the problem. It turns out there is a wide
class of many-particle reaction amplitudes for which such a problem
would be quite meaningful. We have shown that these amplitudes should
be understood as amplitudes of true $n$-particle interaction or
$n$-body forces amplitudes; see details in \cite{17,18}. Here we
reproduce our results taking a line stated in previous section.

The scattering amplitude of the $n$-body reaction may be considered
as a function of the invariant variable $s=(p_1+p_2+\cdots+p_n)^2$
and two unit vectors ${\bf e}$ and ${\bf e}'$ on $(D-1)$-dimensional
sphere $S_{D-1}$, which characterise the initial and final states of
$n$-particle system: 
\[
{\cal F}_n(s;p'_1 p'_2\cdots p'_n,p_1 p_2\cdots p_n)={\cal F}_n
(s;{\bf e}',\bf e).
\]
Dimensionality $D$ of multidimensional space is related to the number
of particles $n$ by the equation $D=3n-3$. There are many ways to
introduce the spherical coordinates in multidimensional space.
Moreover, there are some peculiarities related to a parametrization
of relativistic $n$-particle system. However, we will not concern
this subject here because it does not play any role for our main
goal. For the details we refer to \cite{18} and references therein.

As above we may wright the partial wave expansion
\[
{\cal F}_n(s;{\bf e}',{\bf e})={\cal F}_n(s;\cos\omega)=
\]
\be
=\frac{1}{\pi A_n(s)}\sum_{lm}Y _{lm}
({\bf e}')f_{l}(s)\stackrel{*}{Y}_{lm}({\bf e})
=\frac{1}{\pi
\Gamma_n(s)}\sum_{l}(\frac{l}{\nu}+1)f_l(s)C_l^{\nu}(\cos\omega),
\label{19}
\ee
where $A_n(s)=\Gamma _n(s)/S_{D-1}$, $\Gamma _n(s)$ is $n$-particle
phase space volume, $S_{D-1}=2\pi^{D/2}/\Gamma(D/2)$ is a surface of
$(D-1)$-dimensional
unit sphere, $\cos\omega={\bf e}'\cdot{\bf e}$, and we  have used in
second line of Eq. (\ref{19}) an addition theorem for the
(hyper)spherical harmonics in multidimensional space
\[
\sum_{m=1}^{M(l,\nu)}Y _{lm}({\bf e}')\stackrel{*}{Y}_{lm}({\bf e}) =
\left(\frac{l}{\nu}+1\right)S_{D-1}^{-1}C_l^{\nu}({\bf e}'\cdot{\bf
e}),
\]
\[
\nu = \frac{D}{2}-1,\quad M(l,\nu) =
\frac{(2l+2\nu)\Gamma(l+2\nu)}{\Gamma(l+1)\Gamma(2\nu+1)},
\]
where $C_l^{\nu}(z)$ is Gegenbauer polynomial. Here we contented
ourself with a special class of  $n$-body forces scattering
amplitudes which are invariant under rotation in multidimensional
space (so called $O(D)$-invariant amplitudes).

We will assume that for physical values of the variable $s$ imaginary
part of $n$-body forces scattering  amplitude is analytic function of
$\cos\omega$, regular inside an ellipse $E_n(s)$ in
$\cos\omega$-plane
with center at the origin and with semi-major axis
\be
z_n(s) = 1 + \epsilon_n(s), \quad \epsilon_n(s) = \frac{M_n^2}{2{\bf
Q}^2},\label{20}
\ee
and for any $\cos\omega\in E_n(s)$ is polynomially bounded in the
variable $s$, $M_n$ is some constant of mass dimensionality
independent of $s$, {\bf Q} is global momentum (dependent of $s$) of
$n$-particle system which will be defined later on. Such analyticity
of $n$-body forces scattering amplitudes was called {\it global}
\cite{18}. If it is so, one can wright Cauchy representation for
imaginary part of  $n$-body interaction amplitude
\[
Im {\cal F}_n(s;\cos\omega)=\frac{1}{2\pi i}\oint_{C_n} dz \frac{Im
{\cal F}_n(s;z)}{z - \cos\omega},
\]
where contour $C_n$ is a boundary of an ellipse $E_n(s)$ in
$\cos\omega$-plane with semi-major axis given by Eq. (\ref{20}).
There is a standard generalization of  Heine's expansion of the
Cauchy denominator \cite{10}
\be
\frac{1}{z - t} = \exp (-i\pi\nu)2^{2\nu}[\Gamma(\nu)]^2(z^2-1)^{\nu
-1/2}
\sum_{l=0}^{\infty}(l+\nu)\frac{\Gamma(l+1)}{\Gamma(l+2\nu)}D_l^{\nu}
(z)C_l^{\nu}(t),\label{21}
\ee
which converges absolutely for
\be
|[t+(t^2-1)^{1/2}]/[z+(z^2-1)^{1/2}]|<1.\label{converge}
\ee
In Eq.~(\ref{21}) $D_l^{\nu}(z)$ is a second solution to Gegenbauer's
equation. The restriction (\ref{converge}) requires that the point
$t$ lie within that ellipse in the complex $t$-plane with foci at
$\pm 1$ which passes through the point $t=z$. In particular from
Eq.~(\ref{21}) it follows
\[
D_l^{\nu}(z) = \exp(i\pi\nu)(z^2-1)^{-\nu+1/2}
\frac{1}{2\pi}\int_{-1}^{1} dt
\frac{(1-t^2)^{\nu-1/2}C_n^{\nu}(t)}{z-t}.
\]
As a result we obtain
\be
Im f_l(s) =
\frac{\exp(-i\pi\nu)\nu2^{2\nu}[\Gamma(\nu)]^2\Gamma(l+1)}{\Gamma(l+2
\nu)}\cdot \frac{\Gamma_n(s)}{2 i}\oint_{C_n} dz
(z^2-1)^{\nu-1/2}D_l^{\nu}(z)Im
{\cal F}_n(s;z).\label{22}
\ee
Representation (\ref{22}) is very useful to study an asymptotic
behaviour of partial waves at large global orbital
momentum. Taking into account asymptotic properties of the Gegenbauer
functions $D_l^{\nu}$ \cite{10}
\[
D_l^{\nu}(z)\simeq \frac{\exp(i\pi\nu)l^{\nu-1}}{2^{\nu}
\Gamma(\nu)}(z^2-1)^{-\nu/2}(z +
\sqrt{z^2-1})^{-l-\nu},\quad
|l|\rightarrow\infty, \, |\arg l| < \pi,\, z\in C_n,
\]
and polynomial boundedness $$\max_{z\in C_n}|Im {\cal F}_n(s;z)|\leq
P_n(s),$$ 
$P_n(s)$ is some polynomial in $s$, we find
\[
Im f_l(s)\leq
\Gamma_n(s)P_n(s)\frac{l^{\nu-1}\nu2^{\nu+1}\Gamma(\nu)\Gamma(l+1)}{\
\Gamma(l+2\nu)}\times
\]
\be
\left(\frac{z_n(s)
+\sqrt{z_n^2(s)-1}}{\sqrt{z_n^2(s)-1}}\right)^{1-\nu}
\left(z_n(s)+\sqrt{z_n^2(s)-1}\right)^{-l}, \ \,
|l|\rightarrow\infty.
\label{23}
\ee
Finally if we put $z_n(s) = 1 + \epsilon_n(s)$, $\epsilon_n(s) <<
1,\,
s\rightarrow \infty$ then we get at large values of $s$
\be
Im f_l(s)\leq
\frac{P_n(s,\nu)}{l^{\nu}}\exp\left(-l\sqrt{2\epsilon_n(s)}\right),
\quad
s\rightarrow
\infty,\label{24}
\ee
where
\be
P_n(s,\nu) =
\nu2^{\nu+1}\Gamma(\nu)[2\epsilon_n(s)]^{(\nu-1)/2}\Gamma_n(s
)P_n(s).\label{25}
\ee
Estimate (\ref{25}) shows that partial waves as functions of global
orbital momentum $l$ exponentially decrease at large values of $l$,
i.e. the significant contribution to the partial wave expansion
(\ref{19}) is resulted from partial waves for which the global
orbital momentum does not exceed the quantity
\be
\Lambda =
\left[\frac{\ln
P_n(s,\nu)}{\sqrt{2\epsilon_n(s)}}\right].\label{26}
\ee
The contribution of partial waves with $l>\Lambda$ to the partial
wave expansion will be exponentially small. So, we decompose the
partial wave expansion in two terms
\be
Im {\cal F}_n(s;\cos\omega=1) =
\frac{1}{\pi\Gamma_n(s)}\sum_{l=0}^{\Lambda}(\frac{l}{\nu}+1)Im
f_l(s)C_l^{\nu}(1)
+ Im {\cal F}_n^{\Lambda}(s),\label{27}
\ee
where the second term in Eq. (\ref{27}) contains the contribution of
partial waves with $l>\Lambda$. Taking into account the unitarity
condition (\ref{12}) for the partial waves we get for the first term
in Eq.~(\ref{27}) an estimate 
\[
\frac{1}{\pi\Gamma_n(s)}\sum_{l=0}^{\Lambda}(\frac{l}{\nu}+1)Im
f_l(s)C_l^{\nu}(1)\leq
\frac{1}{\pi\Gamma_n(s)}\sum_{l=0}^{\Lambda}(\frac{l}{\nu}+1)C_l^{\nu
}(1) =
\]
\be
\frac{(2\Lambda+2\nu+1)\Gamma(\Lambda+2\nu+1)}{\pi\Gamma_n(s)\Gamma(2
\nu+2)\Gamma(\Lambda+1)} =
\frac{2\Lambda^{2\nu+1}}{\pi\Gamma_n(s)\Gamma(2\nu+2)}\left(1+O(\frac
{1}{\Lambda})\right),\label{28}
\ee
where we inserted
$C_l^{\nu}(1)=\Gamma(l+2\nu)/[\Gamma(2\nu)\Gamma(l+1)]$. It can
easily
be seen that the second term in Eq.~(27) is
asymptotically small compared to the first one at large values of
$s$, so that we finally get
\be
Im {\cal F}_n(s;\cos\omega=1) <
\frac{2\left[\ln
P_n(s,\nu)\right]^{D-1}}{\pi\Gamma(D)\Gamma_n(s)[2\epsilon_n(s)]^{(
D
-
1)/2}}.\label{29}
\ee
where we have used expression (\ref{26}) for $\Lambda$ and relation
$2\nu=D-2$. By analogy with Eq.~(\ref{16}) let us introduce the
effective radius $R_n(s)$ of $n$-body forces 
\be
R_n(s)\stackrel{def}{=}\frac{\Lambda}{|\bf Q|}=\frac{1}{M_n}\ln
P_n(s,\nu),\label{30}
\ee
where the definition (\ref{26}) of the quantity $\Lambda$ and
expression
(\ref{20}) for $\epsilon_n(s)$ have been used. Now upper bound
(\ref{29}) in terms of such defined quantity $R_n(s)$ takes the form
\be
Im {\cal F}_n(s;\cos\omega=1) <
\frac{2\left[R_n(s)\right]^{D-1}}{\pi\Gamma(D)\Gamma_n(s)[2\epsilon_n
(s)/M_
n^2]^{(D-1)/2}}=J_n(s)S_{D-1}[R_n(s)]^{D-1},\label{31}
\ee
where
\be
J_n(s) = \frac{2}{\pi\Gamma(D)S_{D-1}^2
A_n(s)[2\epsilon_n(s)/M_n^2]^{(D-1)/2}} = \frac{2|{\bf
Q}|^{D-1}}{\pi\Gamma(D)S_{D-1}^2
A_n(s)}.\label{32}
\ee
With account of the generalized optical theorem relating a total
cross section of $n$-body interaction with imaginary part of $n$-body
forces forward scattering amplitude \cite{18}
\[
\sigma_n^{tot}(s) =
\frac{1}{J_n(s)}Im {\cal F}_n(s;\cos\omega=1),
\]
from estimate (\ref{31}) we obtain an upper bound for the
total cross section of $n$-body interaction 
\be
\sigma_n^{tot}(s) < S_{D-1}[R_n(s)]^{D-1}.\label{33}
\ee
Here again, as it should be, upper bound (\ref{33}) has a quite
clear geometrical meaning: the total cross section of $n$-body
interaction is bounded by the area of a surface of
$(D-1)$-dimensional
sphere whose radius is defined by the effective radius of $n$-body
forces. Again all information about global analyticity of $n$-body
interaction amplitudes is hidden in the physical quantity (\ref{30})
which is the effective radius of $n$-body forces. From
Eqs.~(\ref{25}) and (\ref{29}) it follows that
\[
P_n(s,\nu)\sim c_n\,s^{(3n+3)/4},\quad s\rightarrow\infty.
\]
For the the effective radius of $n$-body forces we find
from Eq.~(\ref{30}) in that case
\be
R_n(s) \sim 
\frac{r_n}{M_n}\ln (s/s_0),\quad r_n=\frac{3n+3}{4},\quad
s\rightarrow\infty.\label{34}
\ee
Upper bounds (\ref{31},\ref{33}) are a direct consequence of global
analyticity of $n$-body forces scattering amplitudes which, in one's
turn, is a direct geometrical generalization of analytic properties
of two-body scattering amplitude strictly proved in axiomatic Quantum
Field Theory. At present we do not know to what extend global
analyticity of $n$-particle scattering amplitudes $(n\geq3)$ is a
consequence of general principles of local Quantum Field Theory. The
validity of such an assumption is obvious to us if we rely on the
physical nature of $n$-body forces: our intuition tells us that true
$n$-body interactions should manifest themselves only in the case
when all the $n$ particles are in a sufficiently limited volume. On
the other hand, from the beginning one may, by definition, consider
the $n$-body forces scattering amplitude to be a globally analytic
part of the total $S$-matrix which may always be singled out from it
\cite{18}.  

At last, we have to give the definition of global momentum $|{\bf
Q}|$ for the relativistic $n$-particle system. In this respect, first
of all, note that momentum ${\bf q}$ for two-particle system has been 
defined in a relativistic covariant way. Under any Lorentz
transformation $\Lambda$ from the restricted Lorentz group
$\Lambda\in{\cal L}_+^{\uparrow}$ momentum ${\bf q}$ is transforming
by Wigner rotation: 
\[
{\bf q}\rightarrow{\bf q'} = R_W{\bf q},\quad R_W = L^{-1}(\Lambda
P_{ab})\Lambda L(P_{ab}),
\]
$L(P_{ab})$ is Lorentz boost. This means that $|{\bf q}|$ defined by
Eq.~(\ref{lambda}) is a Lorentz invariant quantity. Moreover, we
would like to emphasize the following asymptotic properties
\be
{\bf q}^2\simeq\frac{1}{4}s,\,s\rightarrow\infty;\quad {\bf q}^2
\simeq
2\mu_2(\sqrt{s}-M_2),\, \sqrt{s}\rightarrow M_2, \,M_2=m_a+m_b,
\,\mu_2=\frac{m_a m_b}{M_2}.\label{35} 
\ee
The expression of ${\bf q}^2$ given by Eq.~{\ref{lambda}} can be
rewritten in the form
\be
{\bf q}^2 = 16s\left(\Gamma_2(s)/S_2\right)^2 =
16sA_2^2(s).\label{36}
\ee
The definition of global momentum for the relativistic $n$-particle
system should be given such as to save the asymptotic properties
shown by Eqs.~(\ref{35}). Such generalization for any number of
particles looks like
\be
{\bf Q}^2=\gamma_n s^{(n-1)/(3n-5)}A_n^{2/(3n-5)}, \label{37}
\ee
where $\gamma_n$ is dimensionless constant
\be
\gamma_n=2^{2n/(3n-5)}\left(\frac{\mu_n}{M_n}\right)^{(2n-4)/(3n-5)},
\, \mu_n=\left(\frac{\prod_{i=1}^nm_i}{M_n}\right)^{1/(n-1)},
\,
M_n=\sum_{i=1}^n m_i. \label{38}
\ee
From the definition (\ref{37}) we have the following asymptotic
properties:
\be
{\bf Q}^2\simeq a_n^2 s,\quad s\rightarrow\infty,\label{39}
\ee
where $a_n$ is dimensionless constant
\be
a_n^2 =
\left(\frac{\Gamma(3n/2-3/2)}{\pi^{(n-1)/2}(n-1)!(n-2)!}\right)^{2/(3
n-5)}\left(\frac{\mu_n}{M_n}\right)^{(2n-4)/(3n-5)},\label{40}
\ee
for example
\[
a_2^2=\frac{1}{4},\quad a_3^2=\left(\frac{\mu_3}{\pi
M_3}\right)^{1/2},
\quad a_3^{-1}(m_1=m_2=m_3)=2.0100... ,
\]
and
\be
{\bf Q}^2\simeq 2\mu_n(\sqrt{s}-M_n),\quad \sqrt{s}\rightarrow
M_n.\label{41}
\ee

\section{Physical applications and discussion}

Let us come back to Eq.~(\ref{13}) and remind an ancient
Pythagoras theorem stated that the sum of first $N$ odd numbers
beginning from unity is equal exactly to the square of $N$ i.e.
\be
\underbrace{1 + 3 + 5 + 7 + \cdots}_{N} = N^2.\label{Pyth}
\ee
This Pythagoras theorem can easily be proved with the help of the
formula for an arithmetical progression. However, Pythagoras theorem
can be proved without a knowledge of the formula for an arithmetical
progression but using only some remarkable observations in a game
with the numbers. We will not touch here the simplest proof, we would
only like to stress a deep link between the Froissart bound and this
Pythagoras theorem. Of course, to take advantage of this link we
have to learn apart from differential calculus and integral calculus
that:
\begin{itemize}
\item Symmetry properties of the space-time continuum are described
by inhomogeneous Lorentz group or Poincar\'e group. We had
also to know how to construct the unitary representations of this
group as well, as it was made in the fundamental paper of
Wigner \cite{11}. 
\item There is a very deep connexion between general physical
principles such as causality, spectrality, unitarity and analytic
properties of physical scattering amplitudes. The very essence of
this connexion is expressed by brilliant Jost-Lehmann-Dyson
representation which provided the fundamental results of Lehmann.
\item It takes many other attainments and the knowledge acquisitions
as well.
\end{itemize}

There is a generalization of Pythagoras theorem (\ref{Pyth}). Really,
let us consider any polynomial $P_n(x)$ degree of $n$
\[
P_n(x)=c_0(P)+c_1(P)x+c_2(P)x^2+\cdots +c_n(P)x^n,
\]
let $S(N)$ be a sum of the polynomial values when the argument $x$
takes an integer value
\[
S(N)\stackrel{def}{=}\sum_{k=0}^N P_n(k),
\]
then it can be proved that $S(N)$ is also a polynomial $Q_{n+1}(N)$
in $N$ degree of $(n+1)$
\be
S(N)=Q_{n+1}(N),\quad  Q_{n+1}(x)=c_0(Q)+c_1(Q)x+c_2(Q)x^2+\cdots
+c_{n+1}(Q)x^{n+1},\label{Pythag}
\ee
and there is correspondence between $c_n(P)$ and $c_n(Q)$:
$c_{n+1}(Q)=c_n(P)/(n+1),\cdots.$ For example, if $P_4(x)$ is
polynomial of fourth degree then we have
\ber
c_5(Q)=c_4(P)/5,\nonumber \\
c_4(Q)=c_3(P)/4+c_4(P)/2,\nonumber \\
c_3(Q)=c_2(P)/3+c_3(P)/2+c_4(P)/3,\nonumber \\
c_2(Q)=c_1(P)/2+c_2(P)/2+c_3(P)/4,\nonumber \\
c_1(Q)=c_0(P)+c_1(P)/2+c_2(P)/6-c_4(P)/30.\nonumber \\
c_0(Q)=c_0(P).\label{p4}
\eer
We will call that statement as a generalized Pythagoras theorem. It
can easily be seen that usual Pythagoras theorem (\ref{Pyth})
corresponds to $P_1(x)=2x+1$. From Eq.~(\ref{28}) it's clear that the
generalized Froissart theorem is related to the generalized
Pythagoras theorem where $P_{D-2}(x)$ is being used. 

In according with the theory held by Pythagoras the unitarity bounds
(\ref{17}) and (\ref{33}) give form to the Unlimited and therefore
they are Limit; see Introduction.

Recently \cite{19,20,21} a simple theoretical formula describing the
global structure of $pp$ and  $p\bar p$ total cross-sections in the
whole range of energies available up today has been derived by an
application of single-time formalism in QFT and general theorems a
l\`a Froissart. The fit to the experimental data with the formula was
made, and it was shown that there is a very good correspondence of
the theoretical formula to the existing experimental data obtained at
the accelerators. Moreover, it turned out there is a very good
correspondence of the theory to all existing cosmic ray experimental
data as well \cite{21}. The predicted values for  $\sigma^{tot}_{pp}$
obtained from theoretical description of all existing accelerators
data are completely compatible with the values obtained from cosmic
ray experiments. The global structure of (anti)proton-proton
total cross section is shown in Figs. 1-2 extracted from papers
\cite{20,21}.

\begin{figure}[t]
\begin{center}
\begin{picture}(188,200)
\put(-35,0){\includegraphics*[]{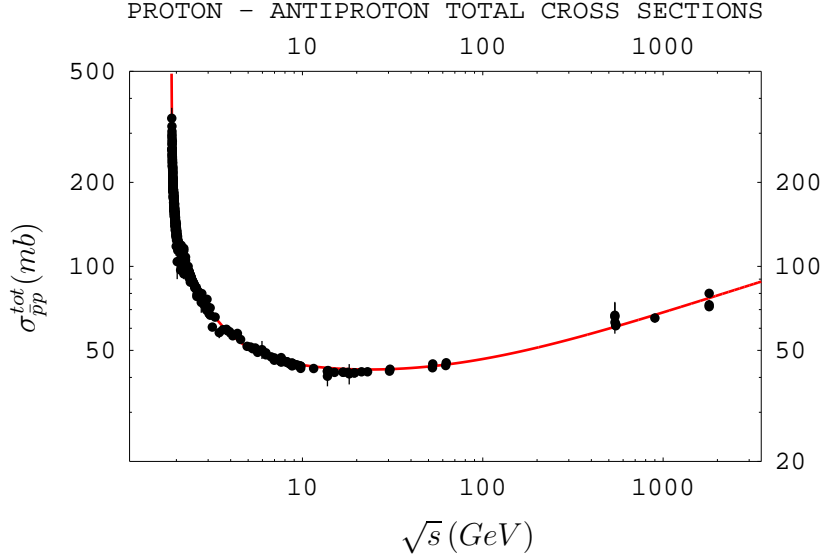}}
\put(92,-15){$\sqrt{s}\, (GeV)$}
\put(-55,65){\rotatebox{90}{$\sigma^{tot}_{\bar{p}p} (mb)$}}
\end{picture}
\end{center}
\vspace*{5mm}
\caption[]{\protect
{The proton-antiproton total cross sections
versus $\sqrt{s}$ compared with the theory. Solid line represents our
fit to the data \cite{19,20}. Statistical and systematic errors added
in quadrature.}}
\label{fig:1}
\end{figure}

\begin{figure}[t]
\begin{center}
\begin{picture}(288,200)
\put(15,10){\includegraphics{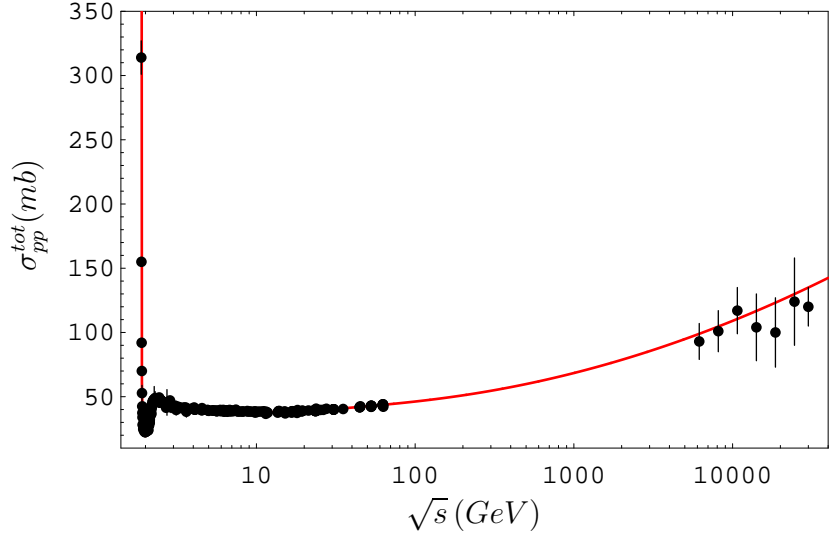}}
\put(144,0){$\sqrt{s}\, (GeV)$}
\put(-5,95){\rotatebox{90}{$\sigma^{tot}_{pp} (mb)$}}
\end{picture}
\end{center}
\caption[]{\protect{The proton-proton total cross-section versus
$\sqrt{s}$ with the cosmic rays data points from Akeno Observatory
and Fly's Eye Collaboration. Solid line corresponds to our theory
predictions \cite{21}.}}
\label{fig:Fig.2}
\end{figure}
The theoretical formula describing the global structure of
(anti)proton-proton total cross section has the following
structure
\be
\sigma_{(\bar p)pp}^{tot}(s) = \sigma^{tot}_{asmpt}(s) 
\left[1 + \chi_{(\bar p)pp}(s)\right],\label{42}
\ee
where
\be
\sigma^{tot}_{asmpt}(s) = 2\pi\left[B_{el}(s) +
(1-\beta)R_3^2(s)\right] = \left[42.0479 + 1.7548
\ln^2(\sqrt{s}/20.74)\right](mb),\label{43}
\ee
\[
B_{el}(s) = R_2^2(s)/2 = \left[11.92 + 0.3036
\ln^2(\sqrt{s}/20.74\right](GeV^{-2}),
\]
\[
R^2_3(s)|_{\beta<<1} = \left[0.40874044 \sigma^{tot}_{asmpt}(s)(mb) -
B_{el}(s)\right](GeV^{-2}) =
\]
\be
=\left[5.267+0.4137\ln^2\sqrt{s}/20.74\right](GeV^{-2}),\label{44}
\ee
\[
\beta = \frac{x^2_{inel}}{4(1+x^2_{inel})},\quad
x^2_{inel}=\frac{R_3^2(s)}{R_d^2}=\frac{2B_{sd}}{R_d^2},
\]
$B_{el}(s)$ is the slope of nucleon-nucleon differential elastic
scattering cross section, $R_2(s)$ is the effective radius of
two-nucleon forces, $R_3(s)$ is the effective radius of three-nucleon
forces, $R_d$  characterizes the internucleon distance in a deuteron,
the functions $\chi_{(\bar p)pp}(s)$ describe low-energy parts of
(anti)proton-proton total cross sections and asymptotically tend to
zero at $s\rightarrow\infty$ (see details in the 
original paper \cite{20}). The mathematical structure of the formula
(\ref{42}) is very simple and physically transparent: the total cross
section is represented in a factorized form. One factor describes
high energy asymptotics of total cross section and it has the
universal energy dependence predicted by the general Froissart
theorem in local Quantum Field Theory. The other factor is
responsible for the behaviour of total cross section at low energies
and it has a complicated resonance structure.  However this factor
has also the universal asymptotics at elastic threshold. It is a
remarkable fact that the low energy part of total cross section has
been derived by application of the generalized Froissart theorem for
a three-body forces scattering amplitude. 

Eq.~(\ref{43}) shows that geometrical scaling in a naive form
$\sigma^{tot}_{asmpt}(s) = \mbox{Const}B_{el}(s)$ is not valid.
However, from  Eq.~(\ref{43}) it follows the generalized  geometrical
scaling which looks like 
\be
\sigma^{tot}_{asmpt}(s) = 2\pi B_{el}(s)[1 +
2\gamma (1-\beta )],\label{geom-scale}
\ee
where $\beta$ is defined above and
\[
\gamma = \frac{R_3^2(s)}{2B_{el}(s)}=\frac{R_3^2(s)}{R_2^2(s)}.
\]

Here, we would like to point out some remarkable features of the
global structure in the (anti)proton-proton total cross sections.
First of all, the (anti)proton-proton total cross sections have a
minimum at $s=s_0$, and the question is what this minimum corresponds
to. It turns out that the effective radius of three-nucleon forces
at the point $s=s_0$ satisfys the following harmonic ratio
\be
\fbox{$\displaystyle R_3(s_0):r_p^{ch}=1:2$}\,,\label{45}
\ee
where $r_p^{ch}=0.88\,fm$ is the proton charge radius. In other
words, at the minimum $s=s_0$ it takes place the {\it``octave
consonance"} of the three-nucleon forces with the proton charge
distribution. 

Going further on, we have applied our approach to study a shadow
dynamics in scattering from deuteron in some details. In this way a
new simple formula for the shadow corrections to the total
cross-section in scattering from deuteron has been derived and new
scaling characteristics with a clear physical interpretation have
been established. We shall briefly sketch the basic results of our
analysis of high-energy particle scattering from deuteron. As has
been shown in \cite{22}, the total cross-section in the scattering
from deuteron can be expressed by the formula
\[
\sigma_{hd}^{tot}(s) = \sigma_{hp}^{tot}(\hat s)
+\sigma_{hn}^{tot}(\hat s) -
\delta\sigma(s), 
\]
where $\sigma_{hd}, \sigma_{hp}, \sigma_{hn}$  are the total
cross-sections in scattering from deuteron, proton and neutron, 
\be
\delta\sigma(s) = \delta\sigma^{el}(s)
+\delta\sigma^{inel}(s)=2\sigma^{el}(s)a^{el}(x_{el}) +
2\sigma_{sd}^{ex}(s)a^{inel}(x_{inel}) ,\label{46}
\ee
\[
\sigma^{el}(s) \equiv \frac{\sigma_{hN}^{tot\,2}(s)}{16\pi
B_{el}(s)},\quad a^{el}(x_{el}) = \frac{x^2_{el}}{1+x^2_{el}},
\quad
x^2_{el} \equiv \frac{2B_{el}(s)}{R_d^2} =
\frac{R_2^2(s)}{R_d^2},
\]
\[
a^{inel}(x_{inel}) = \frac{x^2_{inel}}{(1+x^2_{inel})^{3/2}}, \qquad
x^2_{inel} \equiv \frac{R_3^2(s)}{R_d^2} =
\frac{2B_{sd}(s)}{R_d^2},
\]
the total single diffractive dissociation cross-section
$\sigma_{sd}^{ex}(s)$ is defined by the following equation \cite{22}
\be
\sigma_{sd}^{\varepsilon}(s) =
\pi\int_{M_{min}^2}^{\varepsilon
s}\frac{dM_X^2}{s}\int_{t_{-}(M_X^2)}^{t_{+}
(M_X^2)} dt \frac{d\sigma}{dtdM_X^2}, \label{47}
\ee
where
\be
\varepsilon=\varepsilon^{ex} = \sqrt{2\pi}/2M_N R_d,\label{ex}
\ee
and we supposed that
$\sigma_{hp}^{tot}=\sigma_{hn}^{tot}=\sigma_{hN}^{tot}$ and
$B_{el}^{hp}=B_{el}^{hn}=B_{el}$ at high energies. The first term in
the R.H.S. of Eq.~(\ref{46}) generalizes the known
Glauber correction
\[
\delta\sigma^{el}(s)=\delta\sigma_G(s)=\frac{\sigma_{hN}^{tot\,2}(s)}
{4\pi R_d^2},\quad  x^2_{el}<<1,
\]
but the second term in the R.H.S. of Eq.~(\ref{46}) is totally new
and comes from the contribution of the three-body  forces to the
hadron-deuteron total cross section. 

The expressions for the shadow corrections have quite a 
transparent physical meaning, both the elastic $a^{el}$ and
inelastic $a^{inel}$ scaling functions have a clear physical
interpretation \cite{23}. The function $a^{el}$  measures out a
portion of elastic rescattering events among of all the events during
the  interaction of an incident particle with a deuteron as a whole,
and this function attached to the total probability of elastic
interaction of an incident particle with a separate nucleon in a
deuteron. Correspondingly, the function $a^{inel}$ measures out a
portion of inelastic events of inclusive type among of all the events
during  the interaction of an incident particle with a deuteron as a
whole, and this function attached to the total probability of
single diffraction dissociation  of an incident particle on a
separate nucleon in a deuteron. The scaling variables $x_{el}$ and
$x_{inel}$ have quite a clear physical meaning too. The dimensionless
quantity $x_{el}$ characterizes the effective distances measured in
the units of ``fundamental length", which the deuteron size is, in
elastic interactions, but the similar quantity $x_{inel}$
characterizes the effective distances measured in the units of the
same ``fundamental length" during inelastic interactions.

The functions $a^{el}$ and $a^{inel}$ have a different
behaviour: $a^{el}$ is a monotonic function while $a^{inel}$ has the 
maximum at the point $x^{max}_{inel}=\sqrt{2}$ where 
$a^{inel}(x^{max}_{inel})=2/3\sqrt{3}$. The existence of the maximum
in the function $a^{inel}$ results an interesting physical
effect of weakening the inelastic eclipsing (screening) at
superhigh energies. The energy $s_m$ at the maximum of $a^{inel}$ can
easily be calculated from the equation $R_3^2(s_m)=2 R_d^2$ and
here we faced with the harmonic ratio (in square)
\be
\fbox{$\displaystyle R_3^2(s_m):R_d^2=2:1$}\,.\label{48}
\ee

Using the above mentioned global structure for the
(anti)proton-proton total cross sections, we have made a preliminary
comparison of the new structure for the shadow corrections in elastic
scattering from deuteron with the existing experimental data on
proton-deuteron and antiproton-deuteron total cross sections. The
results of this comparison are shown in Figs.~3-4

\vspace{5mm}
\begin{figure}[hbt]
\begin{center}
\begin{picture}(298,184)
\put(20,10){\includegraphics{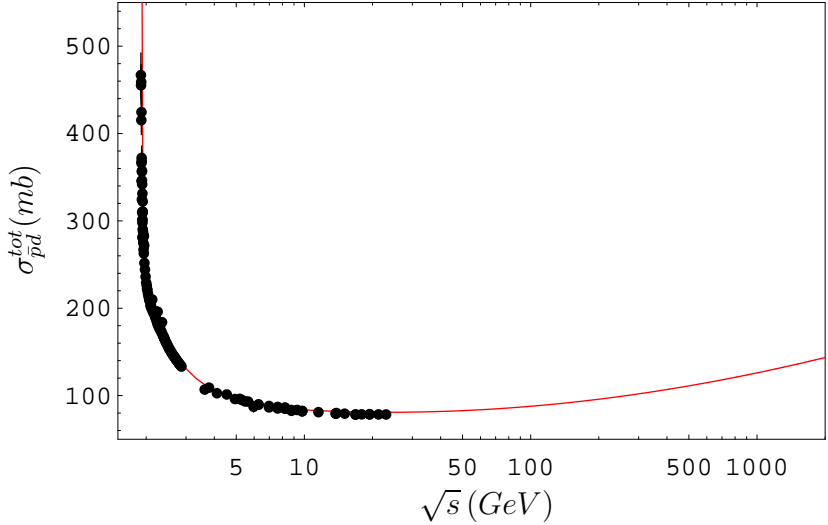}}
\put(155,0){$\sqrt{s}\, (GeV)$}
\put(0,90){\rotatebox{90}{$\sigma^{tot}_{\bar{p}d} (mb)$}}
\end{picture}
\end{center}
\caption{The total antiproton-deuteron cross-section compared with
the theory. Statistical and systematic errors added in
quadrature.}\label{fig3}
\end{figure}

\vspace{5mm}
\begin{figure}[htb]
\begin{center}
\begin{picture}(288,184)
\put(15,10){\includegraphics{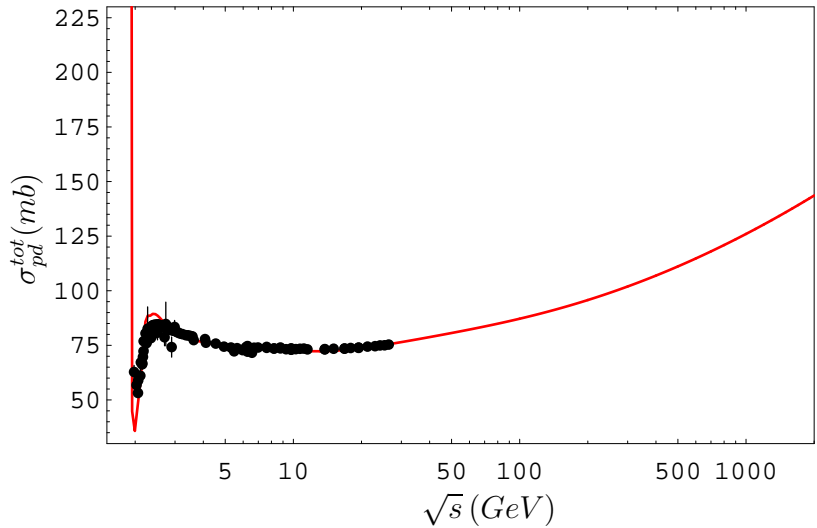}}
\put(155,0){$\sqrt{s}\, (GeV)$}
\put(0,88){\rotatebox{90}{$\sigma^{tot}_{pd} (mb)$}}
\end{picture}
\end{center}
\caption{The total proton-deuteron cross-section compared with the
theory without any free parameters. Statistical and systematic errors
added in quadrature.}\label{fig4}
\end{figure}
We would like to emphasize that in the fit to the data on
antiproton-deuteron total cross sections $R_d^2$ was considered as a
single free fit parameter. After that a comparison with the data on
proton-deuteron total cross sections has been made without any free
parameters: $R_d^2$ was fixed by the previous fit to the data on
antiproton-deuteron total cross sections, and our fit yielded
$R_d^2=66.61\pm 1.16\,GeV^{-2}$. If we take into account the
latest experimental value for the deuteron matter radius
$r_{d,m}=1.963(4)\,fm$ \cite{24} then we can find that the fitted
value for the $R_d^2$ satisfies with a good accuracy the equality
\be
R_d^2 = \frac{2}{3}r^2_{d,m},\quad  (r^2_{d,m} = 3.853\,fm^2 =
98.96\,GeV^{-2}).\label{rd}
\ee
So, we have established a harmonic {\it``consonance"} between  the
internucleon distance in a deuteron and the deuteron matter
distribution
\be
\fbox{$\displaystyle R_d^2:r_{d,m}^2=2:3$}\,.\label{50}
\ee

Now, let us come back to Eq.~(\ref{43}). Taking into account that
$0\leq\beta\leq 1/4$, from the Froissart bound (\ref{17}) and
Eq.~(\ref{43}) we have the following bound
\be
\fbox{$\displaystyle R_3^2(s)< 2R_2^2(s)$}\,.\label{51}
\ee
On the other hand for the effective radii of $n$-body forces we have
obtained an asymptotic behaviour given by Eq.~(\ref{34}) where it
follows from 
\be
\frac{R_3(s)}{R_2(s)}=\frac{4}{3}\cdot\frac{M_2}{M_3},\quad
s\rightarrow\infty.\label{52}
\ee
Bound (\ref{51}) with account of Eq.~(\ref{52}) gives
\be
M_3>\frac{4}{3\sqrt{2}}M_2=\frac{8m_{\pi}}{3\sqrt{2}},\quad
(M_2=2m_\pi).\label{54}
\ee
However, if we conjecture that $M_n=nm_\pi$ which is fulfilled for
$n=2$ then 
\be
\frac{R_3(s)}{R_2(s)}=\frac{8}{9},\quad
s\rightarrow\infty.\label{55}
\ee
The ratio given by Eq.~(\ref{55}) corresponds to the harmonic ratio
for the major second in the major scale in according to Pythagoras
tune; see Introduction. We would like especially to emphasize that
the ratio (\ref{55}) is compatible with the global structure of
(anti)proton-proton(deuteron) total cross sections described above. 

\section{Conclusion}

In this minireview we have tried in the spirit of the Pythagorean
school to show the mathematical, physical and geometrical beauty of
the Froissart theorem. No doubt, we were enchanted with the aesthetic
aspects of the Froissart theorem: there were heard {\it the new notes
of the music of the spheres} produced by the Froissart theorem in the
fundamental dynamics of particles and nuclei. Starting from abstract
mathematical structures of axiomatic Quantum Field Theory by applying
the general theorems, a physically transparent intuitively clear and
visual picture of particles and nuclei interactions was arisen before
our eyes. We found a very simple relations between physically
tangible
quantities which looked like Pythagoras harmonic ratios
mentioned above and hence might be considered as a ``hadronic
symphony" in the fundamental dynamics. In fact, we came back to the
great Pythagorean ideas reformulated in terms of the objects living
in the microcosmos.

It appears that the study of fundamental processes in high energy
elementary particle physics makes it possible to establish a missing
link between cosmos and microcosmos, between the great ancient ideas
and recent investigations in particle and nuclear physics and to
confirm the unity of physical picture of the World. Anyway, we
believe in it.

At last, in our previous papers we repeatedly criticized the so
called supercritical pomeron phenomenology in hadronic physics. In
our opinion this phenomenology might be compared with a ``cacophony"
in particle physics. Certainly, someone likes cacophony in the music.
However, we prefer a symphony in the music and a harmony in the
fundamental dynamics as well.

\section*{Acknowledgements}
It is a great pleasure to thank Professor O.A.~Khrustalev who
initiated, encouraged and supported my scientific flame and research
in the youth and  Professor V.I.~Savrin for friendly and successfully
collaboration in the middle of seventieth.


\begin{thebibliography}{**}
\bibitem{1}
V.V.~Belokurov, O.A.~Khrustalev, O.D.~Timofeevskaya, Quantum
Teleportation -- usual Wonder, in russian, Izhevsk, 2000.
\bibitem{2}
A.V.~Lebedev, Fragments of ancient Greek philosophers, Part I, in
russian, Moscow, ``Nauka", 1989.
\bibitem{3}
M.~Froissart, Phys. Rev. {\bf 123}, 1053 (1961). 
\bibitem{4}
H.~Lehmann, Nuovo Cim. {\bf 10}, 579 (1958).
\bibitem{5}
R.~Jost, H.~Lehmann, Nuovo Cim. {\bf 5}, 1598 (1957).
\bibitem{6}
F.J.~Dyson,  Phys. Rev. {\bf 110}, 1460 (1958).
\bibitem{7}
H.~Lehmann, Nuovo Cimento Suppl. {\bf 14}, 153 (1959).
\bibitem{8}
A.~Martin, Nuovo Cim. {\bf 42A}, 930 (1966); ibid {\bf 44A},
1219 (1966).
\bibitem{9}
G.~Sommer, Nuovo Cim. {\bf 48A}, 92 (1967); {\bf 52A}, 373 (1967);
{\bf 52A}, 850 (1967); {\bf 52A}, 866 (1967).
\bibitem{10}
L.~Durand, P.M.~Fishbane, L.M.~Simmons, Jr., Journ. of Math. Phys.
{\bf 17}, 1933 (1976). 
\bibitem{11}
E.P.~Wigner, {\it On unitary representations of the inhomogeneous
Lorentz group}, Ann. of Math. {\bf 40}, 149 (1939).
\bibitem{12}
O.A.~Khrustalev, V.I.~Savrin, Preprint IHEP 68-19K, Serpukhov, 1968.
\bibitem{13}
O.A.~Khrustalev, A.A.~Logunov, Nguen Van Hieu, in: {\it Problems of
Theoretical Physics}, Essays dedicated to N.N.~Bogolyubov on the
occasion of his Sixtieth Birthday, Publishing House ``Nauka", Moscow,
1969, p.~90.
\bibitem{14}
O.A.~Khrustalev, A.A.~Logunov, A.N.~Tavkhelidze, I.T.~Todorov, Nuovo
Cim. {\bf 30}, 134 (1963).
\bibitem{15}
A.A.~Arkhipov, Sov. J. Theor. Math. Phys. {\bf 74}, 69
(1988); ibid {\bf 83}, 247 (1990).
\bibitem{16}
G.'t Hooft, {\it The Holographic Principle}, e-print hep-th/0003004.
\bibitem{17}
A.A.~Arkhipov, V.I.~Savrin, Sov. J. Theor. Math. Phys. {\bf 49}, 3
(1981).
\bibitem{18}
A.A.~Arkhipov, Rep. on Math. Phys. {\bf 20}, 303 (1984).
\bibitem{19}
A.A.~Arkhipov, {\it What Can we Learn from the Study of Single
Diffractive Dissociation at High Energies?} -- in Proceedings of
VIIIth Blois Workshop on Elastic and Diffractive Scattering,
Protvino, Russia, June 28--July 2, 1999, World Scientific, Singapore,
2000, pp.~109-118; REPORT IHEP 99-43, Protvino, 1999; e-print
hep-ph/9909531. 
\bibitem{20}
A.A.~Arkhipov, {\it On Global Structure of Hadronic Total Cross
Sections}, preprint IHEP 99-45, Protvino, 1999; e-print
hep-ph/9911533.
\bibitem{21}
A.A.~Arkhipov, {\it Proton-Proton Total Cross Sections from the
Window of Cosmic Ray Experiments}, preprint IHEP 2001-23, Protvino,
2001; e-print hep-ph/0108118; in Proceedings of IXth
Blois Workshop on Elastic and Diffractive Scattering, Pruhonice near
Prague, Czech Republic, June 9-15, 2001, eds. V.~Kundrat, P.~Zavada,
Institute of Physics, Prague, Czech Republic, 2002, pp.~293-304.
\bibitem{22}
A.A.~Arkhipov, {\it Three-Body Forces, Single Diffraction
Dissociation and Shadow Corrections in Hadron-Deuteron Total Cross
Sections}, preprint IHEP 2000-59, Protvino, 2000; e-print
hep-ph/0012349; in Proceedings of XVth Workshop on High
Energy Physics and Quantum Field Theory, Tver, Russia, September
7-13, 2000, eds. M.~Dubinin, V.~Savrin,
Institute of Nuclear Physics, Moscow State University, Russia, 2001,
pp.~241-257.
\bibitem{23}
A.A.~Arkhipov,{\it DIFFRACTION 2000: New Scaling Laws in Shadow
Dynamics}, Nucl. Phys. B (Proc. Suppl.) {\bf 99A}, 72 (2001).
\bibitem{24}
F.~Schmidt-Kaler et al., Phys. Rev. Lett. {\bf 70}, 2261 (1993).
\end{thebibliography}
\end{document}